\documentclass[prc,twocolumn,showpacs]{revtex4-1}

\usepackage{enumerate}

\usepackage{booktabs}
\usepackage{color}
\usepackage{graphicx}

\usepackage{amsmath}
\usepackage{amssymb}
\usepackage{times}

\newcommand{\del}[2]{\frac{\partial #1}{\partial #2}}

\newcommand{\bra}{\langle}
\newcommand{\ket}{\rangle}

\newcommand{\beq}{\begin{equation}}
\newcommand{\eeq}{\end{equation}}
\newcommand{\bea}{\begin{eqnarray}}
\newcommand{\eea}{\end{eqnarray}}

\def\fun#1#2{\lower3.6pt\vbox{\baselineskip0pt\lineskip.9pt
 \ialign{$\mathsurround=0pt#1\hfil##\hfil$\crcr#2\crcr\sim\crcr}}}

\begin{document}

\title{
  Classification of resonances and pairing effects on $NA$-scattering within the HFB framework
}

\author{K. Mizuyama$^{1,2}$, H. Cong Quang$^{3,4}$, T. Dieu Thuy$^{3}$, T. V. Nhan Hao$^{3,4}$}
\email{corresponding author: tvnhao@hueuni.edu.vn}

\affiliation{
  \textsuperscript{1}
  Institute of Research and Development, Duy Tan University,
  Da Nang 550000, VietNam
  \\
  \textsuperscript{2}
  Faculty of Natural Sciences,  Duy Tan University, Da Nang 550000, VietNam
  \\
  \textsuperscript{3}
  Faculty of Physics, University of Education, Hue University,
  34 Le Loi Street, Hue City, Vietnam
  \\
  \textsuperscript{4} Center for Theoretical and Computational Physics,
  College of Education, Hue University, 34 Le Loi Street, Hue City, Vietnam
}

\date{\today}

\begin{abstract}
  We analyze the properties of the scattering solutions obtained as
  the pole of the S- and K-matrix with the help of the Jost function
  framework and the Strum-Liouville theory within the
  Hartree-Fock-Bogoliubov(HFB) framework, and clarify the scattering
  solutions which can be defined as the physical state. We found that
  there are three types of the resonances; ``{\it shape resonance}'',
  ``{\it particle-type}'' and ``{\it hole-type quasiparticle resonances}'',
  and another two types of
  solutions are given as the independent S-matrix and K-matrix poles.  
  The shape resonance is formed by the Hartree-Fock(HF) mean field potential,
  is not affected by the pairing correlation so much. The particle-type
  and hole-type quasiparticle resonances originate from the particle and
  hole states by the configuration mixing effect by pairing.
  All of resonance are represented by the S-matrix pole which has the
  corresponding K-matrix pole.
  Two other types of solutions are given by the independent S-matrix and
  K-matrix poles. These poles are formed by the HF mean field potential.
  The effect of pairing for the independent S-matrix pole is small,
  but the one for the independent K-matrix pole has the remarkable effect.
  The independent K-matrix pole destroys the quasiparticle resonance as it
  approaches to the resonance by the pairing effect.
  The wave function of all resonances have the characteristic structure of
  the metastable property. However, the metastable structure of the wave
  function of the quasiparticle resonance can be broken by the independent
  standing wave solution or the Fano effect.
\end{abstract}

\maketitle

\section{Introduction}

Many sharp resonance peaks at the low energy region of the neutron elastic
cross section is one of the most important characteristic of nuclei.
Those peaks have been analyzed by the resonance formula derived from the
R-matrix theory~\cite{rmatrix}. Resonance parameters have been used for the
widely purposes, such as the nuclear power technology, radiation
therapy and so on. 
However, it is difficult to understand the physics of resonance from
the resonance parameters because the R-matrix theory is phenomenological
theory, there is no clear interpretation of physics for each parameters. 
In addition, there are several types of the resonance formula which has
the different interpretation in physics~\cite{rmatrix1,rmatrix2},
although all of those are consistent with the Feshbach projection theory
~\cite{feshbach}. 
As a consequence of the R-matrix and Feshbach projection theory, the coupling
of channels has the crucial role of the production of the sharp resonance peaks.
Recently, the continuum particle-vibration coupling method succeeded to
reproduce the some of sharp resonances, and it was shown that those peaks
originate from the coupling between an in-coming neutron and the collective
excitation (especially the giant resonances) of target nucleus~\cite{mizuyama}. 

The Jost function~\cite{jost} may be the appropriate method to understand the
physics of the resonance because the Jost function is calculated by the wave
function and potential to represent the appropriate boundary condition
for a regular solution of the Schrodinger equation so as to connect the regular
and irregular solutions, and the zeros of the Jost function on the complex
energy plane represent the poles of the S-matrix corresponding to the bound
states and resonances.
However, the single channel has been supposed in the original Jost function.
The extension of the Jost function is necessary so as to take into account
the channels coupling. As a first step of the extension of the Jost function,
we have extended the Jost function within the HFB formalism~\cite{jost-hfb}.
In a broad sense, the HFB formalism is also the channel coupling formalism
for two channels, because the pairing correlation causes the mixing of the
particle and hole configurations.

The pairing correlation is one of the most important correlation to describe
not only the fundamental properties but also the many of varieties of
interesting phenomena of open-shell nuclei.
The roles of the pairing correlation have been discussed for a long time
on the ground state and excited states of open-shell nuclei.
The important roles of pairing correlation for the exotic structure and dynamics
of the neutron-rich nuclei were also revealed such as the two neutron
halo~\cite{2nhalo,2nhalo2},
di-neutron~\cite{dhalo}, anti-halo effect~\cite{antihalo,antihalo2,antihalo3},
pair rotation~\cite{pairrot}
and so on,
in the last decades. Observation of those phenomena by nuclear reaction is
one of the most important issue. The importance of the pairing correlation in 
the two neutron transfer reaction has been discussed~\cite{2ntransfer}.
Very recently, it was shown that the quasiparticle resonance may be possible to be found
as a sharp peak of the neutron elastic scattering cross section off the
open-shell nucleus within the Hartree-Fock-Bogoliubov (HFB) theory~\cite{kobayashi,jost-hfb}. 

In Figs.9 and 11 of \cite{jost-hfb}, we have shown the trajectories of the S-matrix poles 
to show the dependence on the mean value of the pairing for the stable and neutron-rich
unstable nuclei. The S-matrix poles have been classified into two types; the
``hole-type'' and ``particle-type'' poles of quasiparticle resonances. 
The quasiparticle resonance originates from the HF single particle or hole state due to
the particle and hole configuration mixing by the pairing correlation.
The pairing effect on the ``hole-type'' resonance and the interference effect
with continuum have been discussed in terms of the {\it Fano effect}~\cite{jost-fano}.

The shape resonance~\cite{shaperes} is one of the well-known type resonance.
As is written in many textbook of the quantum physics, a shape resonance is
a metastable state in which a nucleon is trapped due to the shape of the
centrifugal barrier of the mean field potential of the target nucleus.
The wave function of the internal region is connected to the one of the
external region through the tunneling effect. The shape resonance is not
affected by the occupation of the nucleons in the potential ({\it i.e.}
no dependence on the Fermi energy). The overall shape of the neutron elastic
cross section is mainly determined by those shape resonances.

At first, we thought all of S-matrix poles which are found within the HFB framework
can be classified in ``particle-type'' or ``hole-type''. 
However, in \cite{jost-hfb}, we noticed that some poles have different type of the dependence
on the pairing from neither ``particle-type'' nor ``hole-type'' poles. This may imply
that we need further classification for the S-matrix poles.
In this paper, we shall, therefore,  try to classify the S-matrix poles and discuss
the pairing dependence on each types of poles. 

In order to discuss the classification of resonances, firstly we shall review the
scattering theory which relevant with the discussion points of this paper in Sec.\ref{theory}.
In Sec.\ref{numericalres}, we discuss the classification of resonances and properties based on the
numerical results. Finally, we draw the sumarry of this paper in Sec.\ref{summary}. 

\section{Theory}
\label{theory}

In the scattering system of the physics, there are three kinds of quantities, the S-,
T- and K-matrix~\cite{newton,leonard}. As is well known, the S-matrix is the unitary matrix which connects
asymptotically sets of the free particle states in the Hilbert space of the physical
states, such as the bound states, virtual states and resonances. The physical states
are represented as the poles of the S-matrix on the complex energy plane.
The T-matrix is a quantity which is directly connected with the cross section.
The K-matrix is defined as the Hermite matrix. All of those quantities are related
each other, and it has been believed that all of those quantities have the same
information but different mathematical properties.

However, it is possible to show that the K-matrix has the poles on the real axis of
the energy, and those poles are the Sturm-Liouville eigen values.
In ~\cite{sasakawa}, the resonances were defined by the Sturm-Lioville eigen values,
but the poles of the S-matrix and K-matrix are slightly different. We shall explain it in this
section. 

\subsection{K-matrix and Sturm-Liouville eigen value}

The purpose of this subsection is to show that the K-matrix poles are the Sturm-Liouville
eigen value. For simplicity, we start the discussion within the Hartree-Fock (HF) framework
ignoring the pairing. The pairing effect is discussed in the latter section.

Within the HF framework, the Lippmann-Schwinger equation is given by
\begin{eqnarray}
  \psi_{0,lj}^{(+)}(r;\epsilon)
  &=&
  F_{l}(kr)
  \nonumber\\
  &&
  +
  \int_0^\infty dr'
  G_{F,l}^{(+)}(r,r';k)
  U_{lj}(r')
  \psi_{0,lj}^{(+)}(r';\epsilon)
  \nonumber\\
  \label{LSHF}
\end{eqnarray}
with the HF potential $U_{lj}$, 
where $\epsilon$ is the incident energy ($\epsilon=\frac{\hbar^2k^2}{2m}$) 
and $G_{F,lj}(r,r';\epsilon)$ is the free particle Green's function defined by
\begin{eqnarray}
  G_{F,l}^{(\pm)}(r,r';k)
  &\equiv&
  \mp i\frac{2mk}{\hbar^2}
  \left[
    \theta(r-r')F_l(kr')O^{(\pm)}_l(kr)
    \right.
    \nonumber\\
    &&
    +
    \left.
    \theta(r'-r)F_l(kr)O^{(\pm)}_l(kr')
    \right].
\end{eqnarray}
With the scaled riccati-spherical Bessel functions 
defined by $F_{l}(kr)=rj_l(kr)$ and $O^{(\pm)}_{l}(kr)=rh_l^{(\pm)}(kr)$.

It should be noted that $\psi_{0,lj}^{(+)}(r;\epsilon)$ is defined by
\begin{eqnarray}
  \psi_{0,lj}^{(+)}(r;\epsilon)
  =
  \frac{u_{lj}(r;\epsilon)}{J_{0,lj}^{(+)}(\epsilon)},
\end{eqnarray}
where $u_{lj}(r;\epsilon)$ is the regular solution of the HF equation
for a given energy $\epsilon$.

$J_{0,lj}^{(+)}(\epsilon)$ is the HF Jost function.
There are three types of the expressions of the Jost function as following.
\begin{eqnarray}
  J_{0,lj}^{(\pm)}(\epsilon)
  &=&
  1\mp \frac{k}{i}\frac{2m}{\hbar^2}
  \int dr O_{l}^{(\pm)}U_{lj}(r)u_{lj}(r;\epsilon)
  \label{jost01}
  \\
  &=&
  1\mp \frac{k}{i}\frac{2m}{\hbar^2}
  \int dr F_{l}^{(\pm)}U_{lj}(r)v^{(\pm)}_{lj}(r;\epsilon)
  \label{jost02}
  \\
  &=&
  \pm\frac{k}{i}
  W_{lj}(u,v^{(\pm)}).
  \label{jost03}
\end{eqnarray}
where $v^{(\pm)}_{lj}(r;\epsilon)$ is the HF irregular solution
which satisfies the asymptotic boundary condition
$\lim_{r\to\infty}v^{(\pm)}_{lj}(r;\epsilon)\to O^{(\pm)}_{l}(kr)$, and
$W_{lj}(u,v^{(\pm)})$ is the Wronskian.

The S-matrix is defined as
\begin{eqnarray}
  S_{lj}^{(0)}(\epsilon)
  \equiv
  \frac{J_{0,lj}^{(-)}(\epsilon)}{J_{0,lj}^{(+)}(\epsilon)},
  \label{S0def}
\end{eqnarray}
and the S-matrix pole $\epsilon_R^0$ is given by
\begin{eqnarray}
  J_{0,lj}^{(+)}(\epsilon_R^0)=0.
  \label{condS0pole}
\end{eqnarray}
It is, therefore, clear that Eq.(\ref{LSHF}) can be rewritten as
\begin{eqnarray}
  u_{lj}(r;\epsilon_R^0)
  &=&
  \int_0^\infty dr'
  G_{F,l}^{(+)}(r,r';k_R)
  U_{lj}(r')
  u_{lj}(r';\epsilon_R^0)
  \label{LSHF-ER}
\end{eqnarray}
for $\epsilon=\epsilon_R^0$.

By introducing the standing wave Green's function $PG_{F,l}$ defined by
\begin{eqnarray}
  G_{F,l}^{(\pm)}(r,r';k)
  &=&
  PG_{F,l}(r,r';k)
  \nonumber\\
  &&
  \mp
  i
  \frac{2mk}{\hbar^2}
  F_l(kr)F_l(kr')
  \label{GF-PGF},
\end{eqnarray}
Eq.(\ref{LSHF}) can be rewritten as
\begin{eqnarray}
  \psi_{0,lj}^{(+)}(r;\epsilon)
  &=&
  \left(
  1-iT^{(0)}_{lj}(\epsilon)
  \right)
  F_{l}(kr)
  \nonumber\\
  &&
  +
  \int_0^\infty dr'
  PG_{F,l}(r,r';k)
  U_{lj}(r')
  \psi_{0,lj}^{(+)}(r';\epsilon)
  \nonumber\\
  \label{LSHF2}
\end{eqnarray}
where $T^{(0)}_{lj}$ is the T-matrix within the HF framework which can be
calculated by
\begin{eqnarray}
  T^{(0)}_{lj}(\epsilon)
  &=&
  \frac{2mk}{\hbar^2}
  \int_0^\infty dr
  F_{l}(kr)
  U_{lj}(r)
  \psi^{(+)}_{0,lj}(r;\epsilon)
  \label{tmat1}
\end{eqnarray}
The K-matrix $K_{lj}^{(0)}$ is expressed by the T-matrix $T^{(0)}_{lj}$ as
\begin{eqnarray}
  K_{lj}^{(0)}(\epsilon)
  &=&
  \frac{T^{(0)}_{lj}(\epsilon)}{1-iT^{(0)}_{lj}(\epsilon)}
  \label{K0-T0},
\end{eqnarray}
and the standing wave function $\psi_{0,lj}^{(S)}(r;\epsilon)$ is defined by
\begin{eqnarray}
  \psi_{0,lj}^{(S)}(r;\epsilon)
  &\equiv&
  \left(1+iK_{lj}^{(0)}(\epsilon)\right)
  \psi_{0,lj}^{(+)}(r;\epsilon)
  \label{psiS-psi}
  \\
  &\equiv&
  \frac{1}{1-iT_{lj}^{(0)}(\epsilon)}
  \psi_{0,lj}^{(+)}(r;\epsilon)
  \label{psiS-psi2}
\end{eqnarray}
Eq.(\ref{LSHF2}) can be rewritten as
\begin{eqnarray}
  \psi_{0,lj}^{(S)}(r;\epsilon)
  &=&
  F_{l}(kr)
  \nonumber\\
  &&
  +
  \int_0^\infty dr'
  PG_{F,l}(r,r';k)
  U_{lj}(r')
  \psi_{0,lj}^{(S)}(r';\epsilon).
  \nonumber\\
  \label{LSHF3}
\end{eqnarray}
If there is an energy $\epsilon_n^0$ which is given by
\begin{eqnarray}
  1-iT^{(0)}_{lj}(\epsilon_n^0)=0,
  \label{condK0pole1}
\end{eqnarray}
then Eq.(\ref{LSHF3}) can be rewritten as
\begin{eqnarray}
  \psi_{0,lj}^{(+)}(r;\epsilon_n^0)
  &=&
  \int_0^\infty dr'
  PG_{F,l}(r,r';k_n^0)
  U_{lj}(r')
  \psi_{0,lj}^{(+)}(r';\epsilon_n^0).
  \nonumber\\
  \label{LSHF3-en}
\end{eqnarray}
using Eq.(\ref{psiS-psi2}), also $\epsilon_n^0$ is a pole of the K-matrix
by Eq.(\ref{K0-T0}).
Eq.(\ref{condK0pole1}) can be represented by using the Jost function as
\begin{eqnarray}
  J_{0,lj}^{(+)}(\epsilon_n^0)+J_{0,lj}^{(-)}(\epsilon_n^0)=0.
  \label{condK0pole2}
\end{eqnarray}
because the K-matrix can be represented by using the Jost function as
\begin{eqnarray}
  K_{lj}^{(0)}(\epsilon)
  =
  i\left(
  \frac{J_{0,lj}^{(-)}(\epsilon)-J_{0,lj}^{(+)}(\epsilon)}
       {J_{0,lj}^{(-)}(\epsilon)+J_{0,lj}^{(+)}(\epsilon)}%
       \right)
       \label{K0-Jost0}
\end{eqnarray}
It can be proved that the Jost function has a symmetric property
$J_{0,lj}^{(-)}(\epsilon)=J_{0,lj}^{(+)*}(\epsilon^*)$.
This requires $\epsilon_n^0$ to be a real number. 

The Sturm-Liouville theory is the theory of the second-order
differential equations of the form:
\begin{eqnarray}
  \del{}{r}
  \left[
    p(r)
    \del{\hat{\phi}(r)}{r}
    \right]
  +q(r)\hat{\phi}(r)
  =
  -\nu w(r)\hat{\phi}(r)
  \label{SLequ},
\end{eqnarray}
where $p(r)$, $q(r)$ and $w(r)$ are positive definite coefficient functions.
For the HF equation, the coefficient functions are given by
\begin{eqnarray}
  p(r)&=&\frac{\hbar^2}{2m}
  \\
  q(r)&=&\epsilon-\frac{\hbar^2l(l+1)}{2mr^2}
  \\
  w(r)&=&-U_{lj}(r),
\end{eqnarray}
when $\epsilon>0$. 
$\nu$ is called the eigen value of the Sturm-Liouville equation.
According to the Sturm-Liouville theory,
there are infinite number of real eigenvalues
$\nu_n$ which can be numbered so that 
$\nu_1< \nu_2< \cdots \nu_\infty$
(Mercer's theorem~\cite{Mercer}). An eigen function $\hat{\phi}_{n,lj}(r)$
has $n-1$ nodes inside the potential.
Note that the orthogonality of $\hat{\phi}_{n,lj}$ is given by
\begin{eqnarray}
  \int_0^\infty dr
  \hat{\phi}_{m,lj}(r)
  w(r)
  \hat{\phi}_{n,lj}(r)
  =\delta_{mn}.
\end{eqnarray}
Completeness is given by
\begin{eqnarray}
  &&
  \sum_{n=1}^\infty
  w(r)
  \hat{\phi}_{n,lj}(r)
  \hat{\phi}_{n,lj}(r')
  \nonumber\\
  &&=
  \sum_{n=1}^\infty
  \hat{\phi}_{n,lj}(r)
  \hat{\phi}_{n,lj}(r')
  w(r')
  \nonumber\\
  &&=
  \delta(r-r').
\end{eqnarray}
The eigensolution $\hat{\phi}_{n,lj}$ satisfies
\begin{eqnarray}
  &&
  \hat{\phi}_{n,lj}(r)
  \nonumber\\
  &&=
  \int_0^\infty dr'
  PG_{F,l}(r,r';k)
  \nu_n U_{lj}(r')
  \hat{\phi}_{n,lj}(r')
  \label{STeq}.
\end{eqnarray}

If we suppose that there is an eigen value which satisfies 
$\nu_n(\epsilon=\epsilon^0_n)=1$, 
it is very easy to notice that Eq.(\ref{LSHF3-en})
and Eq.(\ref{STeq}) are equivalent.

Therefore, the K-matrix pole has the physical meaning as
the standing wave eigen solution of the Sturm-Liouville eigen
value problem, and it is slightly different from the S-matrix
pole. Because the condition for S-matrix pole(Eq.(\ref{condS0pole}))
is not required to have the condition for the K-matrix pole
(Eq.(\ref{condK0pole2})). Eqs.(\ref{condS0pole}) and (\ref{condK0pole2})
are independent each other. This implies that the S-matrix pole and K-matrix pole
may have the different properties.

\subsection{S-, T- and K-matrix within the HFB framework}

In order to discuss the paring dependence of poles of
the S- and K-matrix, it is necessary to introduce the
definition of the S- and K-matrix which are described
within the HFB framework. In this section, we shall derive
the definition of the K-matrix with the help of the HFB
Jost function. 

By using the HFB Jost function, the S-matrix is expressed as
\begin{eqnarray}
  &&
  S_{lj}(E)
  =
  \sum_{s=1,2}
  \left(
  \mathcal{J}_{lj}^{(+)}(E)
  \right)^{-1}_{1s}
  \left(\mathcal{J}_{lj}^{(-)}(E)\right)_{s1}
  \nonumber\\
  &&=
  \frac{
  (\mathcal{J}_{lj}^{(+)}(E))_{22}
  (\mathcal{J}_{lj}^{(-)}(E))_{11}
  -
  (\mathcal{J}_{lj}^{(+)}(E))_{12}
  (\mathcal{J}_{lj}^{(-)}(E))_{21}
  }
  {
  (\mathcal{J}_{lj}^{(+)}(E))_{22}
  (\mathcal{J}_{lj}^{(+)}(E))_{11}
  -
  (\mathcal{J}_{lj}^{(+)}(E))_{12}
  (\mathcal{J}_{lj}^{(+)}(E))_{21}
  }
  \nonumber\\
  &&=
  \frac{\det\mathcal{J}_{lj}^{(+)*}(E^*)}{\det\mathcal{J}_{lj}^{(+)}(E)}
  \label{Smat}
\end{eqnarray}
where $E$ is the complex quasiparticle energy. The incident energy
$\epsilon$ and the quasiparticle energy $E$ are related as
\begin{eqnarray}
  \epsilon=\lambda+E
\end{eqnarray}
where $\lambda (<0)$ is the chemical potential (Fermi energy) given
in the HFB framework.

The denominator of Eq.(\ref{Smat}) is the determinant of the HFB Jost function,
{\it i.e.}
\begin{eqnarray}
  &&
  \det \mathcal{J}_{lj}^{(+)}(E)
  \nonumber\\
  &&
  =
  (\mathcal{J}_{lj}^{(+)}(E))_{22}
  (\mathcal{J}_{lj}^{(+)}(E))_{11}
  -
  (\mathcal{J}_{lj}^{(+)}(E))_{12}
  (\mathcal{J}_{lj}^{(+)}(E))_{21}.
  \nonumber\\
\end{eqnarray}
and the numerator is also expressed as
\begin{eqnarray}
  &&
  \det \mathcal{J}_{lj}^{(+)*}(E^{*})
  \nonumber\\
  &&=
  (\mathcal{J}_{lj}^{(+)}(E))_{22}
  (\mathcal{J}_{lj}^{(-)}(E))_{11}
  -
  (\mathcal{J}_{lj}^{(+)}(E))_{12}
  (\mathcal{J}_{lj}^{(-)}(E))_{21}
  \nonumber\\
\end{eqnarray}
due to the symmetric properties of the HFB Jost function given by
\begin{eqnarray}
  (\mathcal{J}_{lj}^{(\pm)*}(E^*))_{11}
  &=&
  (\mathcal{J}_{lj}^{(\mp)}(E))_{11}
  \\
  (\mathcal{J}_{lj}^{(\pm)*}(E^*))_{12}
  &=&
  (-)^l
  (\mathcal{J}_{lj}^{(\pm)}(E))_{12}
  \\
  (\mathcal{J}_{lj}^{(\pm)*}(E^*))_{21}
  &=&
  (-)^l
  (\mathcal{J}_{lj}^{(\mp)}(E))_{21}
  \\
  (\mathcal{J}_{lj}^{(\pm)*}(E^*))_{22}
  &=&
  (\mathcal{J}_{lj}^{(\pm)}(E))_{22}
\end{eqnarray}
The S-matrix pole $E_R$ is determined by
\begin{eqnarray}
  \det \mathcal{J}_{lj}^{(+)}(E_R)=0.
  \label{Spole}
\end{eqnarray}

The T-matrix is given by
\begin{eqnarray}
  T_{lj}(E)
  &=&
  \frac{i}{2}
  \left(
  S_{lj}(E)-1
  \right)
  \\
  &=&
  \frac{i}{2}
  \left(
  \frac{\det \mathcal{J}_{lj}^{(+)*}(E^*)-\det \mathcal{J}_{lj}^{(+)}(E)}
       {\det \mathcal{J}_{lj}^{(+)}(E)}
  \right),  
  \label{Tmat-jost}
\end{eqnarray}
and the K-matrix is
\begin{eqnarray}
  K_{lj}(E)
  &=&
  \frac{T_{lj}(E)}{1-iT_{lj}(E)}
  \label{Kmat}
  \\
  &=&
  i
  \left(
  \frac{\det \mathcal{J}_{lj}^{(+)*}(E^*)-\det \mathcal{J}_{lj}^{(+)}(E)}
       {\det \mathcal{J}_{lj}^{(+)*}(E^*)+\det \mathcal{J}_{lj}^{(+)}(E)}
       \right)
       \label{KmatJost}
\end{eqnarray}
The K-matrix pole $E_n$ is determined by
\begin{eqnarray}
  \det \mathcal{J}_{lj}^{(+)*}(E_n)+\det \mathcal{J}_{lj}^{(+)}(E_n)=0.
  \label{Kpole}
\end{eqnarray}

In order to analyze the pairing effect, the two potential formula may be
useful. Here, we shall derive the two potential formula for the K-matrix. 
As is introduced in \cite{jost-fano}, the S-matrix and T-matrix can be divided into
the HF and pairing parts as
\begin{eqnarray}
  S_{lj}(E)&=&S^{(0)}_{lj}(E)S^{(1)}_{lj}(E)
  \label{SS0S1}
  \\
  T_{lj}(E)
  &=&
  T^{(0)}_{lj}(E)
  +
  T^{(1)}_{lj}(E)S^{(0)}_{lj}(E)
  \label{TT0T1}
\end{eqnarray}
By inserting Eq.(\ref{TT0T1}) into Eq.(\ref{Kmat}),
we obtain
\begin{eqnarray}
  K_{lj}(E)
  &=&
  \frac{K^{(0)}_{lj}(E)+K^{(1)}_{lj}(E)}{1-K^{(0)}_{lj}(E)K^{(1)}_{lj}(E)}
  \label{K-K0K1}
\end{eqnarray}
with
\begin{eqnarray}
  K^{(0)}_{lj}(E)
  &=&
  \frac{T^{(0)}_{lj}(E)}{1-iT^{(0)}_{lj}(E)}
  \label{K0def}
  \\
  K^{(1)}_{lj}(E)
  &=&
  \frac{T^{(1)}_{lj}(E)}{1-iT^{(1)}_{lj}(E)}.
  \label{K1def}
\end{eqnarray}

\begin{figure}[htbp]
\includegraphics[scale=0.44,angle=0]{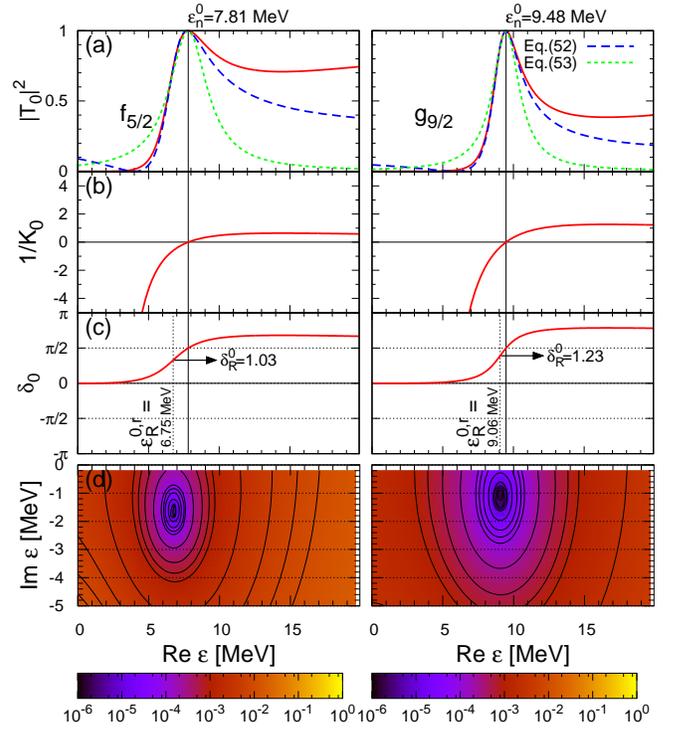}
\caption{(Color online) The numerical results for $f_{5/2}$ and $g_{9/2}$
  of square of T-matrix $|T^{0}_{lj}|^2$,
  the inverse of the K-matrix $1/K^{(0)}_{lj}$,
  and the phase shift $\delta_0$ determined
  $S^{(0)}_{lj}=e^{2i\delta_0}$ are plotted as a function of the
  incident neutron energy $\epsilon$ in the panels (a)-(c),
  respectively. In the bottom panel (d), the square of
  the Jost function $|J_{0,lj}(\epsilon)|^2$ is shown on
  the complex-$\epsilon$ plane.}
\label{fig1}
\end{figure}

\section{Numerical results}
\label{numericalres}

In this paper, we adopt the same Woods-Saxon parameters
for the numerical calculation as in \cite{jost-hfb}.
The chemical potential $\lambda=-8.0$ MeV for the stable
nucleus and $\lambda=-1.0$ MeV for the unstable nucleus
are adopted unless specifically mentioned.

\begin{figure}[htbp]
\includegraphics[scale=0.44,angle=0]{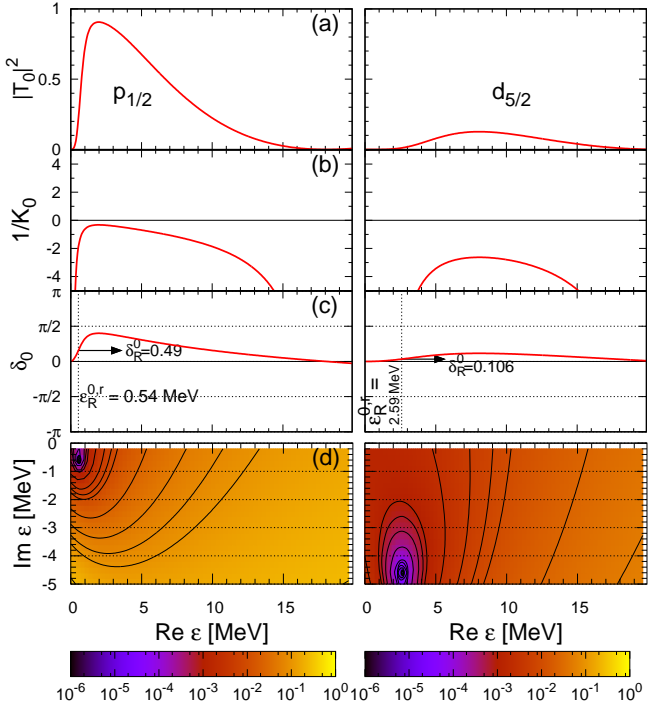}
\caption{(Color online) The same as Fig.\ref{fig1} but for $p_{1/2}$
  and $d_{5/2}$.}
\label{fig2}
\end{figure}

\subsection{Classification of scattering solutions without pairing}
As shown in \cite{jost-hfb}, we can find that the partial cross
section for $s_{1/2}$, $p_{1/2}$, $p_{3/2}$, $d_{5/2}$,
$f_{5/2}$ and $g_{9/2}$ have the contribution for the
total cross section of the neutron elastic scattering
at the zero pairing limit ($\bra\Delta\ket=0$ MeV).

\begin{figure}[htbp]
\includegraphics[scale=0.44,angle=0]{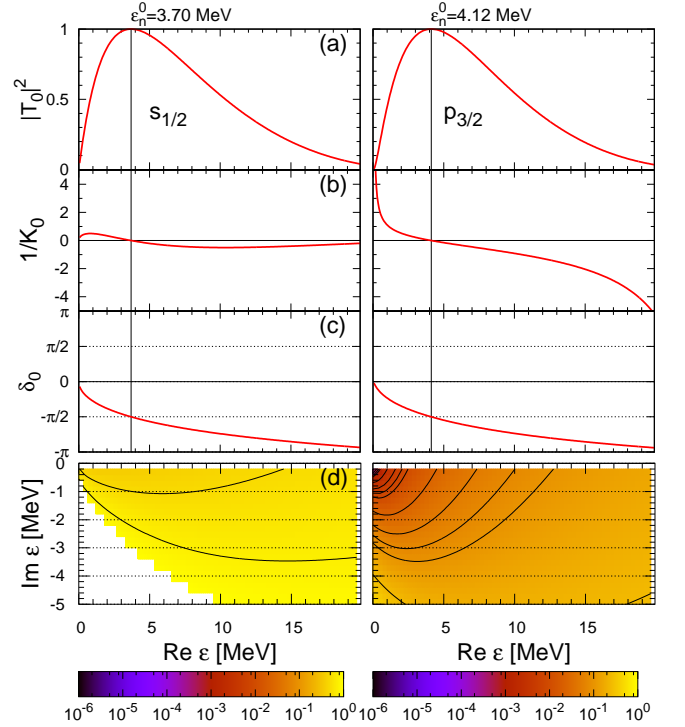}
\caption{(Color online) The same as Fig.\ref{fig1} but for $s_{1/2}$
  and $p_{3/2}$.}
\label{fig3}
\end{figure}

In Figs.\ref{fig1}-\ref{fig3}, we show the square
of the T-matrix $|T^{(0)}_{lj}|^2$ in the panel (a),
this is a quantity which is directly related with the
partial component of the cross section.
The inverse of the K-matrix is shown in the panel (b)
in order to check the existence of the K-matrix pole
on the real axis of the incident energy. In panel (c),
the phase shift which is determined by the S-matrix as
$S_{lj}^{(0)}=e^{2i\delta^0_{lj}}$ is plotted as a function
of the incident neutron energy $\epsilon$.
The square of the Jost function $|J_{0,lj}(\epsilon)|^2$
are shown on the complex-$\epsilon$ plane in order to
show the S-matrix pole. 

In Fig.\ref{fig1}, a peak of $|T^{(0)}_{lj}|^2$ is found
at $\epsilon^0_n=7.81$ MeV for $f_{5/2}$, $\epsilon^0_n=9.48$
MeV for $g_{9/2}$, respectively. The energy of peak is located
at the energy which satisfies $1/K^{(0)}_{lj}(\epsilon^0_n)=0$.
At $\epsilon=\epsilon^0_n$, the phase shift is
obviously $\delta_0=\frac{\pi}{2}$ by definition. 
In the panel (d), we can find the S-matrix pole at
$\epsilon_R^0 = 6.75-i 1.63$ MeV for $f_{5/2}$ and
$\epsilon_R^0 = 9.06-i 1.14$ MeV for $g_{9/2}$, respectively. 
However, $\epsilon_n$ is slightly different from different
from the real part of the S-matrix pole $\epsilon^{0,r}_R$.

When there is a S-matrix pole, $J_{0,lj}^{(+)}(\epsilon)$ is
expressed by the first order Taylor expansion around
$\epsilon=\epsilon_R^0$ as
\begin{eqnarray}
  J_{0,lj}^{(+)}(\epsilon)
  \sim
  (\epsilon-\epsilon_R^0)
  \left.\frac{dJ_{0,lj}^{(+)}(\epsilon)}{d\epsilon}\right|_{\epsilon=\epsilon_R^0}
  \label{approxJ}
\end{eqnarray}
where $\epsilon_R^0=\epsilon_R^{0,r}-i\epsilon_R^{0,i}$.
$\epsilon_R^{0,i}$ is necessary to be small for this approximation. 

The S-matrix is also approximately expressed as
\begin{eqnarray}
  S^{(0)}_{lj}(\epsilon)
  &=&
  \frac{J_{0,lj}^{(-)}(\epsilon)}{J_{0,lj}^{(+)}(\epsilon)}
  \sim
  \frac{(\epsilon-\epsilon_R^{0*})
  \left.\frac{dJ_{0,lj}^{(-)}(\epsilon)}{d\epsilon}\right|_{\epsilon=\epsilon_R^{0*}}
  }{
  (\epsilon-\epsilon_R^0)
  \left.\frac{dJ_{0,lj}^{(+)}(\epsilon)}{d\epsilon}\right|_{\epsilon=\epsilon_R^0}
  }
  \label{Smat-approx}
\end{eqnarray}
Therefore we can obtain
\begin{eqnarray}
  S^{(0)}_{lj}(\epsilon=\epsilon_R^{0,r})
  \sim
  -
  \frac{
  \left.\frac{dJ_{0,lj}^{(-)}(\epsilon)}{d\epsilon}\right|_{\epsilon=\epsilon_R^{0,*}}
  }{
  \left.\frac{dJ_{0,lj}^{(+)}(\epsilon)}{d\epsilon}\right|_{\epsilon=\epsilon_R^0}
  }
  \label{Smat-at-er}
\end{eqnarray}
Since the S-matrix $S^{(0)}_{lj}(\epsilon=\epsilon_R^{0,r})$ is expressed as
$S^{(0)}_{lj}(\epsilon=\epsilon_R^{0,r})=e^{2i\delta_R^0}$ by using the phase-shift
$\delta_R^0$ at $\epsilon=\epsilon_R^{0,r}$, we can obtain
\begin{eqnarray}
  \left.\frac{dJ_{0,lj}^{(-)}(\epsilon)}{d\epsilon}\right|_{\epsilon=\epsilon_R^{0,*}}
  \sim
  -e^{2i\delta_R^0}
  \left.\frac{dJ_{0,lj}^{(+)}(\epsilon)}{d\epsilon}\right|_{\epsilon=\epsilon_R^0}
  \label{Jmapprox}
\end{eqnarray}
Applying Eqs.(\ref{approxJ}) and (\ref{Jmapprox}) to the condition
Eq.(\ref{condK0pole2}), we can obtain
\begin{eqnarray}
  \epsilon_n^0
  &\sim&
  \epsilon_R^{0,r}
  +\epsilon_R^{0,i}
  \cot\delta_R^0
  \label{eneR}
\end{eqnarray}
Using Eqs.(\ref{Smat-approx}) and (\ref{eneR}), the approximated expression
of the T-matrix can be obtained as
\begin{eqnarray}
  &&
  T^{(0)}_{lj}(\epsilon)
  \nonumber\\
  &&=
  \frac{i}{2}
  \left(
  S^{(0)}_{lj}(\epsilon)-1
  \right)
  \nonumber\\
  &&\sim
  \frac{-i}{\epsilon-\epsilon^0_n+\epsilon_R^{0,i}\frac{e^{i\delta_R^0}}
  {\sin\delta_R^0}}
  \left[
    (\epsilon-\epsilon_n^0)e^{i\delta_R^0}\cos\delta_R^0
    +\epsilon_R^{0,i}\frac{e^{i\delta_R^0}}{\sin\delta_R^0}
    \right]
  \nonumber\\
  \label{Tmat-approx}
\end{eqnarray}
If $\delta_R^0$ is supposed to be $\delta_R^0\sim\frac{\pi}{2}$, then
Eq.(\ref{Tmat-approx}) becomes
\begin{eqnarray}
  T^{(0)}_{lj}(\epsilon)
  \sim
  \frac{\epsilon_R^{0,i}}{\epsilon-\epsilon_n^0+i\epsilon_R^{0,i}}
  \label{Tmat-approx2}
\end{eqnarray}
because
\begin{eqnarray}
  \frac{e^{i\delta_R^0}}{\sin\delta_R^0}
  &\sim& i
  \\
  \cos\delta_R^0
  &\sim& 0.
\end{eqnarray}
\begin{figure}[htbp]
\includegraphics[scale=0.44,angle=0]{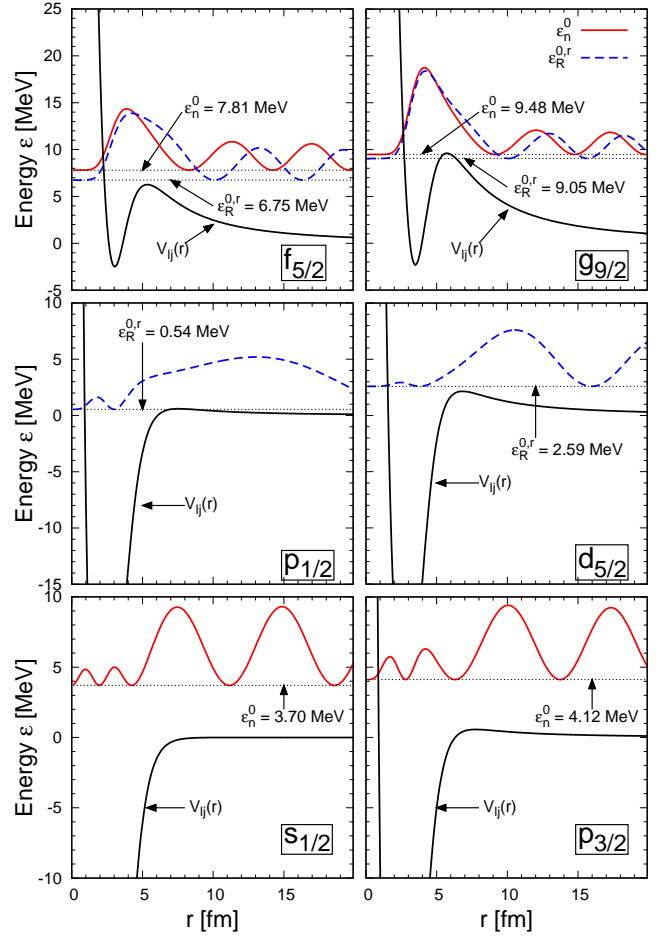}
\caption{(Color online) The square of the scattering wave function $|\psi_{0,lj}^{(+)}|^2$.
  $V_{lj}(r)$ is the potential defined by $V_{lj}(r)=U_{lj}(r)+\hbar^2l(l+1)/2mr^2$. 
  The solid red and dashed blue curves are the square of the wave function
  at $\epsilon=\epsilon_n^0$ and $\epsilon_R^{0,r}$, respectively.}
\label{fig4}
\end{figure}

Eq.(\ref{Tmat-approx2}) is the Breit-Wigner resonance
formula. In the panel (a) of Fig.\ref{fig1}, the square
of T-matrices calculated with Eqs.(\ref{Tmat-approx})
and (\ref{Tmat-approx2}) are plotted by the dashed blue and
dotted green curves, respectively. One can see that
Eq.(\ref{Tmat-approx}) can express the asymmetric shape
of the square of the T-matrix very well. The difference
between the exact result and approximated one is due to
the higher order contribution which are dropped in
Eq.(\ref{Jmapprox}). The higher order contribution seems
to be more important for the description of the shape at
higher energy. The difference between $\epsilon_n^0$ and
$\epsilon_R^{0,r}$ is due to the phase shift $\delta_R^0$
which is determined by the S-matrix at $\epsilon=\epsilon_R^{0,r}$.
The imaginary part of the S-matrix pole $\epsilon_R^{0,i}$
represents the half of the width of the Breit-Wigner type
peak of the square of the T-matrix. 

In Fig.\ref{fig2}, we can find the S-matrix poles at
$\epsilon=0.54-i 0.56$ and $2.59-i 4.63$ MeV for $p_{1/2}$
and $d_{5/2}$, respectively. However, no K-matrix is found
in the panel (b). In the panel (a), we can see the broad
peak shape of square of the T-matrix, but obviously the peak
energy is quite different from $\epsilon_R^{0,r}$, this is
due to the values of the phase shift at $\epsilon=\epsilon_R^{0,r}$
which differ from $\frac{\pi}{2}$.

In Fig.\ref{fig3}, we can find the peak of the square of
the T-matrix at $\epsilon=3.70$ and $4.12$ MeV
for $s_{1/2}$ and $p_{3/2}$, respectively. And we can confirm
that those peaks are poles of the K-matrix from the panel (b).
However, no S-matrix pole is found in the panel (d). Therefore,
it may be possible to interpret that those peaks are the standing
wave eigen solutions which don't have the lifetime. 

\begin{figure}[htbp]
\includegraphics[scale=0.44,angle=0]{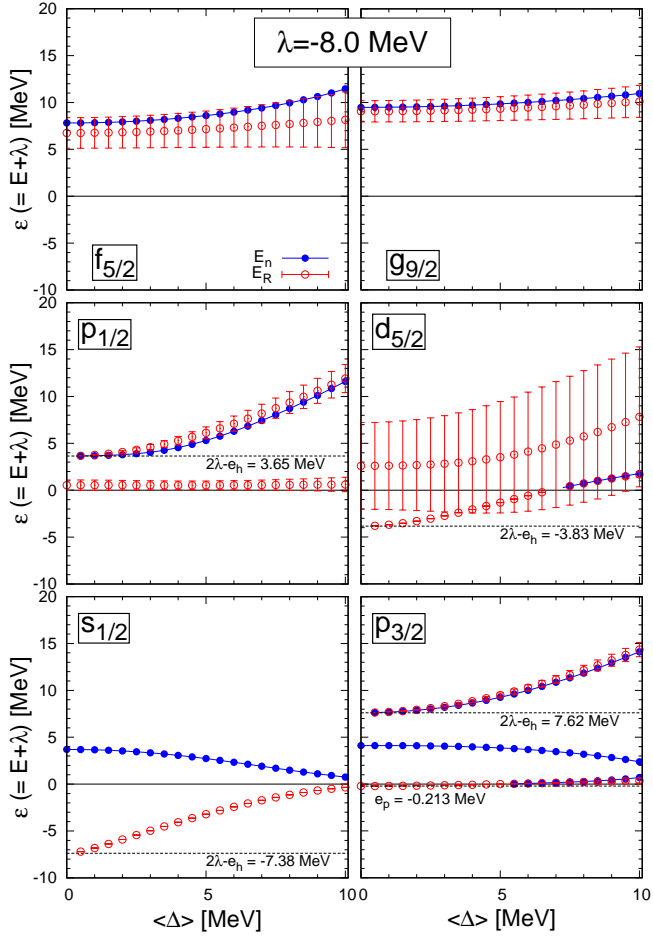}
\caption{(Color online) $E_n$(or $E_R^r$)$+\lambda$ plotted as a function of the mean pairing gap
  $\bra\Delta\ket$ with $\lambda=-8.0$ MeV (stable nucleus). The filled blue circles
  with the solid curve represent the standing solution $E_n$, and the unfilled red
  circles with the error-bars represent $E_R^r$ and $E_R^i$, respectively. }
\label{fig5}
\end{figure}

The square of the wave functions $|\psi_{0,lj}^{(+)}|^2$ are shown
in Fig.\ref{fig4} together with the potential $V_{lj}(r)=U_{lj}(r)+\hbar^2l(l+1)/2mr^2$
(the solid black curve).
For $f_{5/2}$ and $g_{9/2}$, the amplitude at
the internal region of the potential is larger than the one at
the outside of the potential. The wave function which exhibits
the internal structure in the nucleus is connected with the
free particle states outside the nucleus asymptotically. 
This is a behavior which can be interpreted as a metastable state.
Both the S- and K-matrix poles for $f_{5/2}$ and $g_{9/2}$ are
found near the top of the centrifugal barrier of the potential.
These properties exhibit the typical properties of the so-called
``{\it shape resonance}'' written in many textbooks.
It should be noted that the K-matrix pole exists in the half of
the width given by $\epsilon_R^{0,i}$ centered on $\epsilon_R^{0,r}$. 
The K-matrix poles of $s_{1/2}$ and $p_{3/2}$ are found at the
continuum energy region. The behavior of the wave function exhibits
a typical behavior of the continuum states, and the wave length of
the wave function inside of the potential is reflected by the
potential depth. The outer amplitude is larger than the inner one.
The S-matrix poles for $p_{1/2}$ and $d_{5/2}$ are found near the
top of the centrifugal barrier although the height of the barrier
is very low. The amplitude of square of the wave function outside
the potential is much larger than the one inside the potential,
even though the imaginary part of the S-matrix pole for $p_{1/2}$
is small. As one can see in Fig.\ref{fig2}, the phase shift is
very small at $\epsilon=\epsilon_R^{0,r}$ for $p_{1/2}$ and
$d_{5/2}$. This means that the interference between the scattering
wave and outgoing wave without scattering is very small. Namely,
the incident wave can not enter the nucleus at $\epsilon=\epsilon_R^{0,r}$
for $p_{1/2}$ and $d_{5/2}$ even though the centrifugal barrier
is very small. 
\begin{figure}[htbp]
\includegraphics[scale=0.44,angle=0]{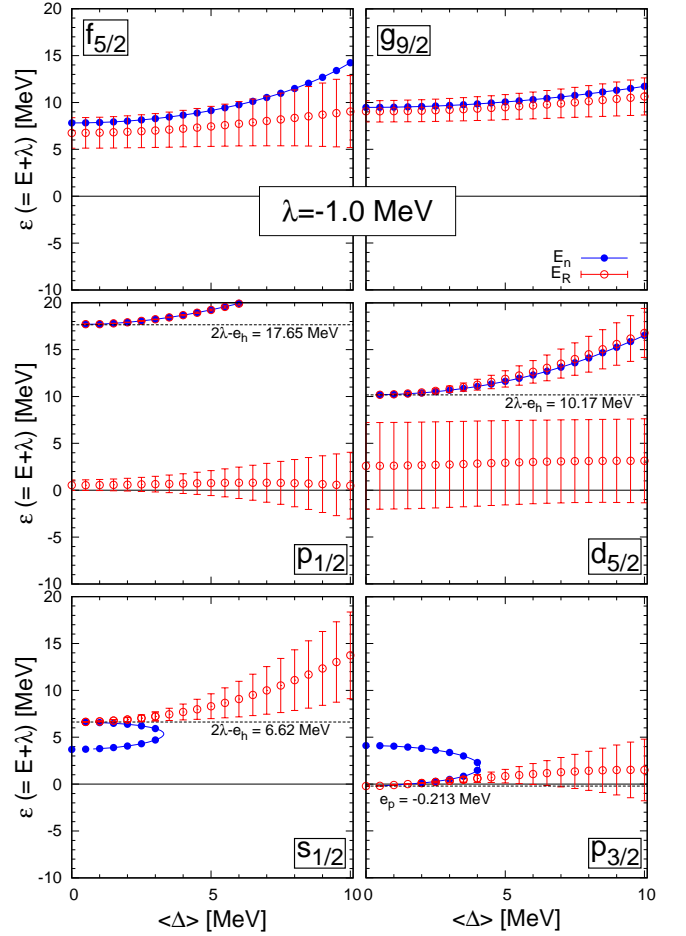}
\caption{(Color online) The same as Fig.\ref{fig5} but $\lambda=-1.0$ MeV (Unstable nucleus).}
\label{fig6}
\end{figure}

From the analysis shown above, basically the S- and K-matrix poles
are independent physical solutions, and the existence of a K-matrix pole 
in the width of the S-matrix pole is necessary in order to form
a resonance which has the property as a metastable state. 

Thereafter, we shall call a S-matrix pole which has the
corresponding the K-matrix pole nearby as a ``resonance''
About other solutions, we shall call a K-matrix pole to which has no
corresponding S-matrix pole as a ``standing wave solution'', and a S-matrix
pole to which has no corresponding the K-matrix pole as just a
(independent) ``S-matrix pole''. 

In the next section, we discuss the pairing dependence of these three types
of scattering solutions including the quasiparticle resonance. 

\subsection{Pairing effect on scattering solutions}
\begin{figure}[htbp]
\includegraphics[scale=0.44,angle=0]{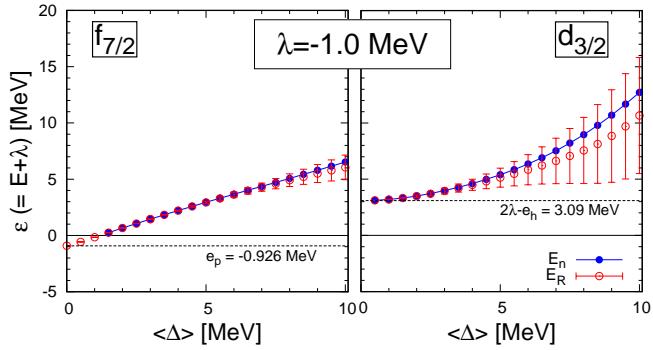}
\caption{(Color online) The same as Fig.\ref{fig5} but for $f_{7/2}$ and $d_{3/2}$ with
  $\lambda=-1.0$ MeV (Unstable nucleus).}
\label{fig7}
\end{figure}

In Figs.\ref{fig5}, \ref{fig6} and \ref{fig7}, 
standing wave solutions $E_n$ and S-matrix poles $E_R (=E_R^r-iE_R^i)$ are plotted
as a function of the mean pairing gap $\bra\Delta\ket$ within the range
$0\leq\bra\Delta\ket\leq 10$ MeV. 
The $E_n$ and $E_R$ are obtained as the numerical solution of 
Eqs.(\ref{Spole}) and (\ref{Kpole}), respectively. 
The standing wave eigen solutions(K-matrix poles) $E_n$ are shown by the filled
blue circles which are connected by solid curve, the S-matrix poles
$E_R (=E_R^r-iE_R^i)$ are shown by the unfilled red circles (for $E_R^r$) with
the error-bars(for $E_R^i$). 
The results for the stable target of nucleus($\lambda=-8.0$ MeV) are shown in
Fig.\ref{fig5}. Figs.\ref{fig6} and \ref{fig7} are results for the unstable target
of nucleus ($\lambda=-1.0$ MeV).       

The pairing dependence of the shape resonances for $f_{5/2}$ and
$g_{9/2}$ with $\lambda=-8.0$ MeV (stable nucleus) and $\lambda=-1.0$ MeV
(unstable nucleus) are shown at the top panels in Figs.\ref{fig5} and
\ref{fig6}, respectively. It is very easy to notice that the standing wave
solution is more sensitive for the pairing than the S-matrix pole in the
sense of the energy shift. The standing wave solution $E_n$ is shifted to
the higher energy as the pairing gap increases. The pairing gap has a
tendency to makes $E_R^i$ (the imaginary part of the S-matrix pole $E_R$)
larger, but the effect is small, $E_R^r$ is stable for the variation of
the pairing gap. 
The shape resonance is a resonance with an $S^{(0)}_{lj}$ pole corresponding
to the $K^{(0)}_{lj}$ pole. The S- and K-matrix are expressed by
Eqs.(\ref{SS0S1}) and (\ref{K-K0K1}) in the HFB framework, respectively.
If Eq.(\ref{Smat-approx}) is good approximation for $S^{(0)}_{lj}$ and
$S^{(0)}_{lj}$ has no pole, then 
it is very clear that the pole of $S_{lj}$ is given by the pole of
$S^{(0)}_{lj}$ from Eq.(\ref{SS0S1}), {\it i.e.} $S^{(1)}_{lj}$ has no
effect on the pole of $S_{lj}$. If the higher order terms of the Taylor
expansion of the Jost function, $S^{(1)}_{lj}$ can be seen as the
effect of the pairing on the pole of $S_{lj}$.
If Eq.(\ref{Smat-approx}) is good approximation for $S^{(0)}_{lj}$,
we can obtain an approximated expression for $T^{(0)}_{lj}$ as given by
Eq.(\ref{Tmat-approx}).
Then the approximated $K^{(0)}_{lj}$ is given by
\begin{eqnarray}
  K^{(0)}_{lj}(E)
  &\sim&
  \frac{1}{E+\lambda-\epsilon_n^0}
  \left(
  \frac{\epsilon_R^{0,i}}{\sin^2\delta^0_R}
  \right)
  +\cot\delta^0_R
  \label{K0approx}.
\end{eqnarray}
The standing wave eigen value $E_n$ is determined by $1/K_{lj}(E_n)=0$,
{\it i.e.}
\begin{eqnarray}
  K^{(1)}_{lj}(E_n)=1/K^{(0)}_{lj}(E_n).
  \label{Encond}
\end{eqnarray}
This is derived from Eq.(\ref{K-K0K1}).

\begin{figure}[htbp]
\includegraphics[scale=0.44,angle=0]{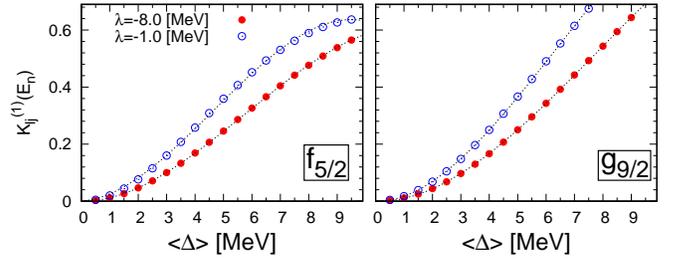}
\caption{(Color online) The pairing gap dependence of $K^{(1)}_{lj}(E_n)$ for
  $f_{5/2}$ and $g_{9/2}$ with $\lambda=-8.0$ and $-1.0$ MeV.
  The dotted curves are the adjusted curves by using the function
  $a \bra\Delta\ket^3+b \bra\Delta\ket^2+c \bra\Delta\ket$.
  The adjusted parameters are shown in Table.\ref{table1} and
  \ref{table2}.}
\label{K1dep}
\end{figure}

By inserting Eq.(\ref{K0approx}) into Eq.(\ref{Encond}), we can obtain
\begin{eqnarray}
  E_n+\lambda\sim \epsilon_n^0
  +
  \frac{K^{(1)}_{lj}(E_n)\epsilon_R^{0,i}}
  {\sin^2\delta^0_R(1-K^{(1)}_{lj}(E_n)\cot\delta^0_R)}
  \label{Enapprox}.
\end{eqnarray}
Since $\delta^0_R\sim\pi/2$ for the shape resonances of
$f_{5/2}$ and $g_{9/2}$ as shown in Fig.\ref{fig1}, Eq.(\ref{Enapprox})
can be expressed as
\begin{eqnarray}
  E_n+\lambda\sim \epsilon_n^0
  +
  K^{(1)}_{lj}(E_n)\epsilon_R^{0,i}
  \label{Enapprox2}.
\end{eqnarray}
This is the reason why $E_n$ of $f_{5/2}$ is more sensitive for the pairing
than the one of $g_{9/2}$.
In Fig.\ref{K1dep}, we show $K^{(1)}_{lj}(E_n)$ as a function of $\bra\Delta\ket$.
The dotted curves are the fitted function curve by using 
$a \bra\Delta\ket^3+b \bra\Delta\ket^2+c \bra\Delta\ket$. The adjusted parameters
for $f_{5/2}$ and $g_{9/2}$ are shown in Table.\ref{table1} and Table.\ref{table2},
respectively.
From the results of adjusted parameters, we can see that $\bra\Delta\ket^2$ term
is dominant to express the pairing gap dependence since $b$ is largest value.
We can see the clear dependence of $b$ on $\lambda$, but no difference between
$f_{5/2}$ and $g_{9/2}$.

As is shown in Fig.\ref{fig2}, the independent S-matrix poles are found for
$p_{1/2}$ and $d_{5/2}$ at the zero pairing limit. The pairing effect on those
S-matrix poles at the middle panels of Figs.\ref{fig5} and \ref{fig6}.
There is almost no effect of the pairing for the S-matrix poles of
$p_{1/2}$ with $\lambda=-8.0$ MeV and $d_{5/2}$ with $\lambda=-1.0$ MeV.
The hole-type quasiparticle resonances are found at higher energy than
the S-matrix poles, except $d_{5/2}$ with $\lambda=-8.0$ MeV.
When the hole-type quasiparticle resonance exists,
$T^{(1)}_{lj}(E)$ is given by
\begin{eqnarray}
  T_{lj}^{(1)}(E)
  &=&
  \frac{\Gamma_{lj}(E)/2}{E-\lambda+e_h-F_{lj}(E)+i\Gamma_{lj}(E)/2}
  \label{T1-holeres},
\end{eqnarray}
where
\begin{eqnarray}
  F_{lj}(E)
  &=&
  \frac{2m}{\hbar^2}
  \frac{2}{\pi}
  P
  \int_0^\infty dk'k'^2
  \frac{|\bra \psi_{0,lj}^{(+)}(k')|\Delta|\phi_{h,lj}\ket|^2}{k^2_1(E)-k'^2},
  \nonumber\\
  \label{defF}
  \\
  \Gamma_{lj}(E)/2
  &=&
  \frac{2m k_1(E)}{\hbar^2}
  |\bra \psi_{0,lj}^{(+)}(k_1(E))|\Delta|\phi_{h,lj}\ket|^2,
  \label{defgm}
\end{eqnarray}
as is derived in \cite{jost-fano}.

$S_{lj}^{(1)}$ and $K_{lj}^{(1)}$ are given by
\begin{eqnarray}
  S_{lj}^{(1)}(E)
  &=&
  \frac{E-\lambda+e_h-F_{lj}(E)-i\Gamma_{lj}(E)/2}
       {E-\lambda+e_h-F_{lj}(E)+i\Gamma_{lj}(E)/2}
  \label{S1-holeres},
  \\
  K_{lj}^{(1)}(E)
  &=&
  \frac{\Gamma_{lj}(E)/2}{E-\lambda+e_h-F_{lj}(E)}
  \label{K1-holeres}.
\end{eqnarray}

By inserting Eq.(\ref{S1-holeres}) into Eq.(\ref{SS0S1}), we obtain
\begin{eqnarray}
  S_{lj}(E)
  &=&
  S_{lj}^{(0)}(E)
  \frac{E-\lambda+e_h-F_{lj}(E)-i\Gamma_{lj}(E)/2}
       {E-\lambda+e_h-F_{lj}(E)+i\Gamma_{lj}(E)/2}
       \nonumber\\
       \label{S-holeres}.
\end{eqnarray}
Therefore, the pole of $S_{lj}^{(0)}(E)$ gives the independent
S-matrix pole, and the S-matrix pole for the hole-type quasiparticle
resonance is
given by
\begin{eqnarray}
  E_R+\lambda=2\lambda-e_h+F_{lj}(E_R)-i\Gamma_{lj}(E_R)/2,
  \label{condER-hole}
\end{eqnarray}

By inserting Eq.(\ref{K1-holeres}) into Eq.(\ref{Encond}), we can obtain
the condition for the standing wave solution $E_n$ of hole-type resonance as
\begin{eqnarray}
  E_n+\lambda=2\lambda-e_h+F_{lj}(E_n)+K_{lj}^{(0)}(E_n)\frac{\Gamma_{lj}(E_n)}{2}
  \nonumber\\
  \label{EqEnholeres}
\end{eqnarray}
From Eqs.(\ref{condER-hole}) and (\ref{EqEnholeres}), we can obtain
\begin{eqnarray}
  E_n=E_R^r+K_{lj}^{(0)}(E_n)\frac{\Gamma_{lj}(E_n)}{2}
  \label{EnERr-hole}
\end{eqnarray}
where
\begin{eqnarray}
  E_R^r=\lambda-e_h+F_{lj}(E_n).
  \label{ERrexp}
\end{eqnarray}
Therefore we can see that 
both $E_R^r$ and $E_n$ increase as the pairing gap increases. 
As we can see from Fig.\ref{fig2}, the sign of $K_{lj}^{(0)}(E_n)$ for $p_{1/2}$ and
$d_{5/2}$ is negative. Eq.(\ref{EnERr-hole}) indicates $E_R^r > E_n$ for the hole-type
quasiparticle resonance of $p_{1/2}$ and $d_{5/2}$ when $K_{lj}^{(0)}(E_n)<0$.
This is consistent with the results of $p_{1/2}$ in Fig.\ref{fig5} and $p_{1/2}$ and
$d_{5/2}$ in Fig.\ref{fig6}. 

In the case of $d_{5/2}$ with $\lambda=-8.0$ MeV, a hole-type quasi-bound state
(this becomes resonance for $\bra\Delta\ket\geq 7.5$ MeV) is found and pushes up the S-matrix
pole as the pairing increases. It seems that a quasi-bound state prevents the
S-matrix pole to stay at the same position. Further investigation may be necessary
to clarify the detail mechanism of this property. 

The same pairing effect can be seen for the hole-type $d_{3/2}$-resonance with
$\lambda=-1.0$ MeV shown in the right panel of Fig.\ref{fig6}, but $E_n>E_R^r$.
This indicates $K_{lj}^{(0)}(E_n)>0$ by Eq.(\ref{EnERr-hole}). 
In Fig.\ref{fig7}, we show the quasiparticle bound state or resonance for the
partial wave components which don't have any visible contribution of $|T|^2$
at the zero pairing limit ({\it i.e.} $|T|^2\to|T_0|^2\simeq 0$ with
$\bra\Delta\ket\to 0$). 
The left panel is the particle-type quasiparticle bound state/resonance for
$f_{7/2}$ which originates from a $f_{7/2}$ particle bound state. We have confirmed
that the particle-type quasiparticle bound states (such as $p_{3/2}$ of
Fig.\ref{fig6} and $f_{7/2}$ of Fig.\ref{fig7}) obey the well-known formula 
$E_R^r=\sqrt{(\epsilon_p-\lambda)^2+\bra\Delta\ket^2}$ for $\bra\Delta\ket\leq 3.0$
MeV, approximately.

As is shown in Fig.\ref{fig3}, the independent K-matrix poles are found for
$s_{1/2}$ and $p_{3/2}$ as the standing wave solution at the zero pairing limit.
The pairing effect for the standing wave solutions are very tricky.
In Fig.\ref{fig5} ($\lambda=-8.0$ MeV), the independent standing wave solutions
$E_n$ for both $s_{1/2}$ and $p_{3/2}$ are shifted to lower energy as the pairing
increases. However, the standing wave solution of $s_{1/2}$ with $\lambda=-1.0$ MeV
is shifted to higher energy as the pairing increases, but a hole-type quasiparticle
resonance is found just above the standing wave solution. The K-matrix pole of the
resonance goes down as the pairing increases, although the S-matrix pole shifts
to higher energy and the width becomes larger. And then, when those two K-matrix
poles meet each other, the K-matrix poles are vanished. Finally, the hole-type
resonance becomes an independent S-matrix pole. A similar thing happens also
for $p_{3/2}$, but the quasiparticle resonance which is found just below the
standing wave solution for $p_{3/2}$ is the particle-type as is shown in the
bottom right panel of Fig.\ref{fig6}. The hole-type resonance for $s_{1/2}$ and
particle-type resonance for $p_{3/2}$ become the independent S-matrix poles for
$\bra\Delta\ket \geq 3.30$ and $\geq 4.08$ MeV, respectively.

\begin{table}
  \caption{The parameters of the fitting function
    $a \bra\Delta\ket^3+b \bra\Delta\ket^2+c \bra\Delta\ket$
    to $K_{lj}^{(1)}(E_n)$ for $f_{5/2}$ shown in the left panel
    of Fig.\ref{K1dep}.}
  \label{table1}
\begin{ruledtabular}
\begin{tabular}{cccc}
$\lambda$ & $a$ & $b$ & $c$ \\
\colrule
$-8.0$ & $-8.2\times 10^{-4}$ & $0.014$ & $-0.0016$ \\
$-1.0$ & $-1.4\times 10^{-3}$ & $0.020$ & $0.0075$\\
\end{tabular}
\end{ruledtabular}
\end{table}

\begin{table}
  \caption{The parameters of the fitting function
    $a \bra\Delta\ket^3+b \bra\Delta\ket^2+c \bra\Delta\ket$
    to $K_{lj}^{(1)}(E_n)$ for $g_{9/2}$ shown in the right
    panel of Fig.\ref{K1dep}.}
  \label{table2}
\begin{ruledtabular}
\begin{tabular}{cccc}
$\lambda$ & $a$ & $b$ & $c$ \\
\colrule
$-8.0$ & $-5.7\times 10^{-4}$ & $0.013$ & $-0.0028$ \\
$-1.0$ & $-1.1\times 10^{-3}$ & $0.020$ & $-0.0015$\\
\end{tabular}
\end{ruledtabular}
\end{table}

The independent K-matrix pole at the zero pairing limit is given by
\begin{eqnarray}
  \frac{1}{K^{(0)}_{lj}(E_n^{(0)})}=0.
\end{eqnarray}
By using the first order approximation of the Taylor expansion,
the inverse of $K^{(0)}_{lj}(E)$ can be expressed as
\begin{eqnarray}
  \frac{1}{K^{(0)}_{lj}(E)}
  \sim
  \frac{E-E_n^{(0)}}{c_{lj}}
  \label{invK0approx}
\end{eqnarray}
where $c_{lj}$ is a real number defined by
\begin{eqnarray}
  \frac{1}{c_{lj}}
  =
  \left(
  \frac{d}{dE}
  \frac{1}{K^{(0)}_{lj}(E)}
  \right)_{E=E_n^{(0)}}.
\end{eqnarray}
By inserting Eqs.(\ref{K1-holeres}) and (\ref{invK0approx}) into
Eq.(\ref{Encond}), we can obtain
\begin{eqnarray}
  (E_n-E_n^{(0)})(E_n-E_R^r)
  =
  c_{lj}\Gamma_{lj}(E_n)/2,
  \label{EqEnholeres2}
\end{eqnarray}
Note that the expression of $E_R^r$ given by Eq.(\ref{ERrexp}) is used.
By supposing that the energy dependence of $F_{lj}$ and $\Gamma_{lj}$
is small, the solutions of Eq.(\ref{EqEnholeres2}) is given by
\begin{eqnarray}
  E_n
  &=&
  \frac{1}{2}
  (E_n^{(0)}+E_R^r)
  \nonumber\\
  &&
  \pm
  \frac{1}{2}
  \sqrt{(E_n^{(0)}+E_R^r)^2-4(E_n^{(0)}E_R^r-c_{lj}\Gamma_{lj}/2)}
  \nonumber\\
  &=&
  \left\{
  \begin{array}{c}
    E_n^{(+)} \\
    E_n^{(-)}
  \end{array}
  \right.
\end{eqnarray}
At the zero pairing limit, $E_R^r\to\lambda-e_h$ and $\Gamma_{lj}\to 0$,
therefore $E_n$ becomes
\begin{eqnarray}
  \lim_{\bra\Delta\ket\to 0}E_n
  \to
  \left\{
  \begin{array}{c}
    E_n^{(0)} \\
    \lambda-e_h
  \end{array}
  \right.
\end{eqnarray}
We can notice that the sign of $c_{lj}$ is negative from Fig.\ref{fig3}.
The pairing dependence is included in $E_R^r$ and $\Gamma_{lj}$, both
values increase as the pairing gap increases. Therefore, a
``{\it critical pairing gap} $\bra\Delta\ket_c$" exists which
satisfies
\begin{eqnarray}
(E_n^{(0)}+E_R^r)^2-4(E_n^{(0)}E_R^r-c_{lj}\Gamma_{lj}/2)=0.
\end{eqnarray}
With this ``{\it critical pairing gap} $\bra\Delta\ket_c$",
$E_n$ is given by $E_n=\frac{1}{2}(E_n^{(0)}+E_R^r)$. 
And $E_n$ doesn't exist for $\bra\Delta\ket>\bra\Delta\ket_c$. 

\begin{figure}[htbp]
\includegraphics[scale=0.44,angle=0]{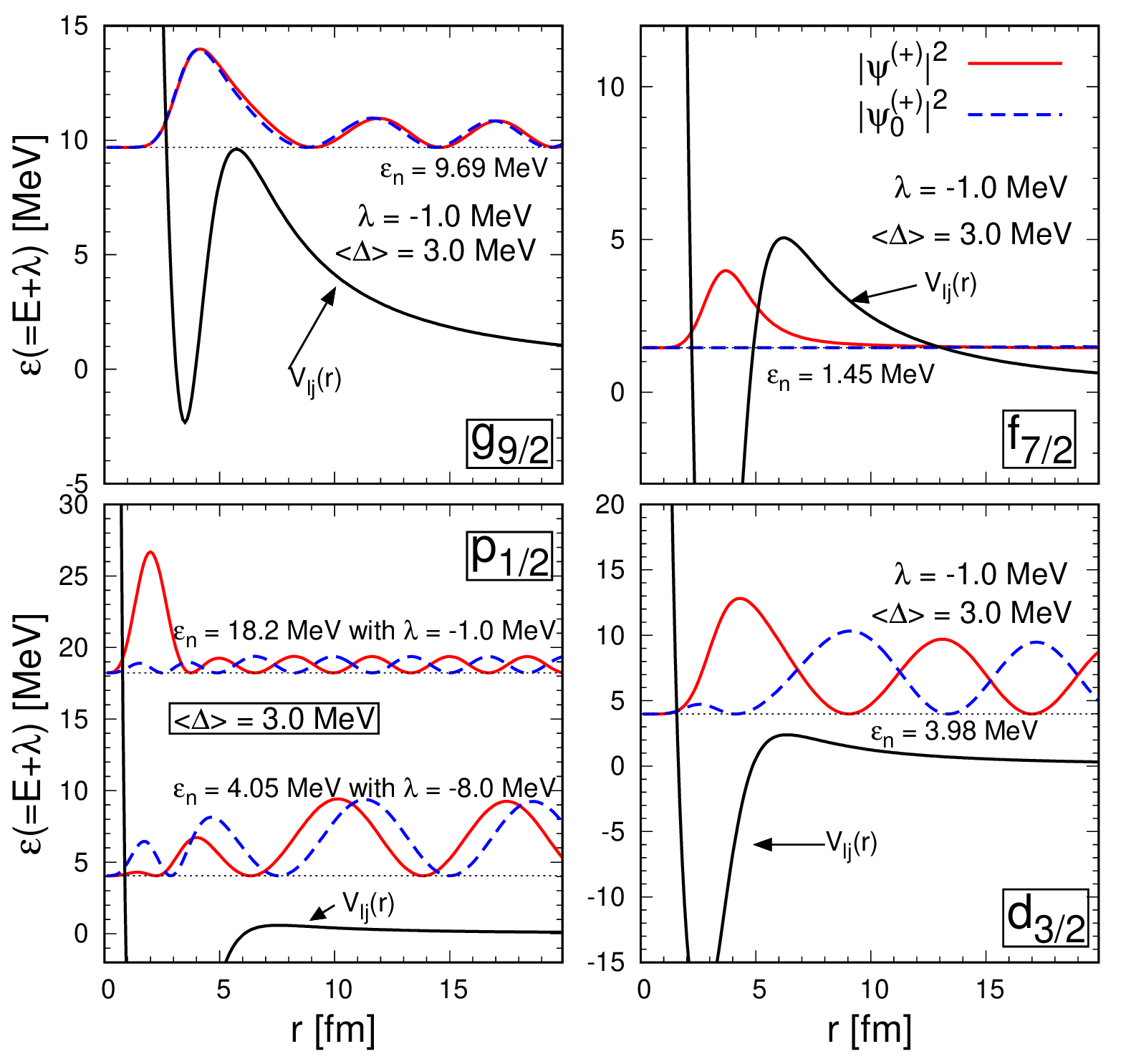}
\caption{(Color online) The square of the scattering wave functions for resonances.
  The upper component of the HFB scattering wave function $|\psi_1^{(+)}|^2$,
  and HF scattering wave function $|\psi_0^{(+)}|^2$ are plotted as a function
  of $r$ [fm] by the red solid and blue dashed curves, respectively. The HF
  potential $V_{lj}=U_{lj}+\hbar^2 l(l+1)/2mr^2$ is plotted together by the black
  solid curve. See text for details.}
\label{psihfb}
\end{figure}

The existence of the ``{\it critical pairing gap} $\bra\Delta\ket_c$"
exhibits the breaking effect of the quasiparticle resonance by the
standing wave solution and pairing. The quasiparticle resonance is
a resonance which is formed by the pairing effect. However, the
quasiparticle resonance can be broken by the pairing effect if
there is an independent standing wave solution nearby. 

\subsection{Pairing effect on the scattering wave function}

In Fig.\ref{psihfb}, the scattering wave functions of the resonances are
plotted as a function of $r$. The red solid curves represent the square of the
upper component of the HFB scattering wave function $|\psi^{(+)}_{1,lj}|^2$,
the blue dashed curves are the HF wave function $|\psi^{(+)}_{0,lj}|^2$.
The difference of those exhibits the effect of the pairing on the scattering wave
function. 

In the upper left panel, the wave function of the shape resonance of $g_{9/2}$
with $\lambda=-1.0$ MeV are shown. The shape of the wave function exhibits the 
metastable property of the state which is reflected to the sharpness of the
peak of the cross section.

The upper right panel shows the wave function of the particle-type quasiparticle
resonance of $f_{7/2}$ with $\lambda=-1.0$ MeV which originates from a HF particle
state $e_p=-0.926$ MeV. The shape of the wave function almost represents the
characteristic of the bound state.
The corresponding peak of the cross section is expected to be a very sharp peak
as a bound state embedded in the continuum. 

\begin{figure}[htbp]
\includegraphics[scale=0.44,angle=0]{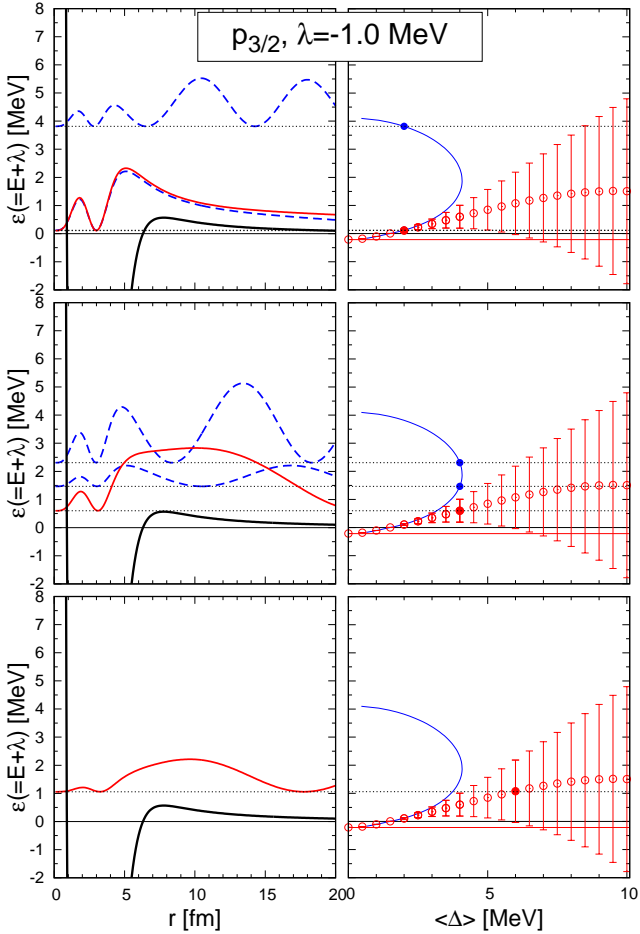}
\caption{(Color online) The scattering wave function for a particle-type quasiparticle
  resonance and independent standing wave solution of $p_{3/2}$ with
  $\lambda=-1.0$ MeV are plotted in the left panels. In the right panels,
  the resonance energy $E_R (=E_R^r-iE_R^i)$ is represented by the red
  circle ($E_R^r$) and error-bar ($E_R^i$), the standing wave solution $E_n$
  is represented by the blue solid curve and circle. Those are plotted as
  a function of the mean pairing gap $\bra\Delta\ket$. The top, middle and
  bottom panels are representing figures for $\bra\Delta\ket=2.0$, $4.0$
  and $6.0$ MeV, respectively.}
\label{psihfb2}
\end{figure}

The wave functions of $p_{1/2}$ hole-type quasiparticle resonance which originates
from a hole state $e_h=-19.65$ MeV is shown in the lower right panel. The wave functions
at $\epsilon=18.2$ MeV and $4.05$ MeV are the wave function with $\lambda=-1.0$ MeV and
$-8.0$ MeV, respectively. The wave function with $\lambda=-1.0$ MeV exhibits the typical
behavior of the metastable state, however, the wave function with $\lambda=-8.0$ MeV
does not have the metastable structure. The $p_{1/2}$ hole-type quasiparticle resonance
is the resonance which has the asymmetric shape of the cross section due to the
``{\it Fano effect}'' caused by the pairing, as is introduced in \cite{jost-fano}.
The Fano effect is the quantum interference effect between the bound state and
continuum. The asymmetry parameter $q$ of the Fano formula is defined by the inverse
of $K^{(0)}_{lj}$. As is known, the shape of the cross section becomes the asymmetric
shape and the Breit-Wigner type of shape with the small and large values of $q$,
respectively. The hole-type quasiparticle resonance for $p_{1/2}$ with $\lambda=-8.0$ MeV 
has the small value of $q$, contrary, the hole-type quasiparticle resonance for
$p_{1/2}$ with $\lambda=-1.0$ MeV has the large value of $q$.
The wave function of the hole-type quasiparticle resonance for $d_{3/2}$ with $\lambda=-1.0$
MeV is shown in the lower right panel. This resonance has been analyzed in \cite{jost-fano} as an
example of the resonance which has large value of $q$. This resonance also has the metastable
structure of the wave function. 
The physical meaning of $q$ is the transition probability to the ``modified quasi-hole''
state at the resonance energy as is introduced in \cite{jost-fano}. Therefore, the large transition
probability due to the ``{\it Fano effect}'' is necessary to form the metastable structure
of the scattering wave function for the hole-type quasiparticle resonance. 

In the last section, we explained that the independent standing wave solution can
break the quasiparticle resonance due to the pairing effect. In Fig.\ref{psihfb2},
we demonstrate how the resonance wave function is broken by the independent standing
wave solution. The wave functions for the resonance (red solid and blue dashed curves)
and standing wave solution (blue dashed curve) for $p_{3/2}$ with $\lambda=-1.0$ MeV
are shown in the left panels. The top, middle and bottom panels are the wave functions
with $\bra\Delta\ket=2.0$, $4.0$ and $6.0$ MeV, respectively.
When $\bra\Delta\ket=2.0$ MeV, the resonance wave function still keeps the metastable structure.
An independent standing wave is found at higher energy. As the pairing increases, the
resonance wave function is separated into the wave functions for the S-matrix pole and
the standing wave solution (K-matrix pole). Tow standing wave solutions approaches and vanishes
at the critical pairing gap $\bra\Delta\ket_c$. The metastable structure of the resonance
wave function is broken and is changed to the wave function structure of the independent
S-matrix pole. 

\section{Summary}
\label{summary}

As is well-known, the S-matrix pole represents a physical state in the scattering
which is so-called ``resonance''. The Jost function formalism is one of the powerful tool
to calculate the S-matrix by using the potential directly. 
In \cite{jost-hfb}, we extended the Jost formalism to take into account the pairing correlation, and 
showed the trajectories of the quasiparticle resonance on the complex
quasi-energy plane varying the pairing gap. Within the HFB framework, it was already
known that there are two types of the quasiparticle resonance, the particle- and hole-type
of the quasiparticle resonances. As is expected, the trajectories of each types of
resonances showed the different trajectories.
However, we found the another types of S-matrix pole which has the different trajectories
from both particle- and hole-type quasiparticle resonances. This fact implies the existence of
the another type of the physical state which has the different effect of the pairing.

In this paper, we clarified the scattering solutions of the $NA$-scattering system within
the HFB framework. For the classification of the scattering solution, firstly, we reviewed
the theory of the S-, T- and K-matrix within the HF framework by ignoring the pairing.
With the help of the Jost function framework and the Strum-Liouville theory, we found that
the poles of the S-matrix and K-matrix are independent. The K-matrix pole is found on the
real axis of the complex energy, and is the standing wave solution which is given as the
eigen solution of the Strum-Liouville eigen value problem for the scattering system.
By analyzing the properties of those scattering solutions, we found that the T-matrix
has the Breit-Wigner type form when there is a corresponding standing wave solution near
the real part of the S-matrix pole, and also the wave function has the metastable
structure. The imaginary part of the S-matrix pole becomes the width of the resonance
peak which is relevant to the life time of the resonance. The independent K-matrix pole
and S-matrix pole can exist. The K-matrix pole is located at the peak of which the square
of the T-matrix becomes one, but the width which is relevant to the life time can not be
defined. The wave function for the K-matrix pole does not have the metastable structure.
The independent S-matrix pole can not form the Wigner type form of the T-matrix even if
the imaginary part of the pole is small, and the wave function at the energy of the real part
of the pole does not have the metastable structure. Even the interference of the wave function
with the in-coming/out-going plane wave is very small. We, therefore, found out that
the ``resonance'' (which has the metastable property) can be formed by the S-matrix pole
which has the corresponding K-matrix pole nearby the real part of the S-matrix pole. 

Secondary, we investigated the classification of the scattering solutions and their
properties especially in terms of the pairing effect within the HFB framework. With
the help of the HFB Jost function framework, we defined and calculated the S-, T- and
K-matrix. In order to analyze the pairing effect, we divided the T-matrix into the HF
term $T^{(0)}$ and the pairing term $T^{(1)}$. The S-matrix and K-matrix are also
expressed by using $S^{(0)}$, $S^{(1)}$, $K^{(0)}$ and $K^{(1)}$ as given by Eqs.(\ref{SS0S1}) and
(\ref{K-K0K1}).
The ``resonance'' is formed by the pole of the S-matrix $S$ which has the corresponding
pole of the K-matrix $K$. If the poles of the S-matrix and K-matrix originate from
$S^{(0)}$ and $K^{(0)}$, the resonance is called the ``{\it shape resonance}''. The pairing
effect is caused mainly by $K^{(1)}$. The effect of pairing on the S-matrix pole is very
small. 
If the poles the S-matrix and K-matrix originate from $S^{(1)}$ and $K^{(1)}$, the resonance
is called the ``{\it quasiparticle resonance}''. If the poles of $S^{(1)}$ and $K^{(1)}$
originate from the particle or hole states, the resonance is called the ``particle-type''
or ``hole-type'' quasiparticle resonance. If there is an independent pole of $K^{(0)}$, the
K-matrix pole of the quasiparticle resonance and the independent pole approach each other as
the pairing gap increases, and both poles are disappeared when two poles come to the same
energy at the critical pairing gap, and then the quasiparticle resonance becomes the independent
S-matrix pole. It seems that the quasiparticle is broken by the standing wave solution
(independent K-matrix pole) due to the pairing effect. Actually, we could see that the
independent standing wave destroy the metastable structure of the wave function of the
quasiparticle resonance by the pairing effect, in the analysis of the wave function.
It was also confirmed that, when the Fano parameter $q$ is small, the ``{\it Fano effect}''
also erases the metastable structure of the hole-type quasiparticle resonance wave function 
by keeping the poles for the resonance as is.

\section{Acknowledgments}
This work is funded by Vietnam National Foundation for Science and Technology Development (NAFOSTED)
under grant number “103.04-2019.329”.


\begin{thebibliography}{00}
\bibitem{rmatrix} A.M.Lane and R.G.Thomas, Rev. Mod. Phys. ~{\bf 30} 257 (1958).
\bibitem{rmatrix1} P.~L.~Kapur and R.~E.~Peierls, Proc.~Roy.~Soc.~{\bf 166A}~277(1938).
\bibitem{rmatrix2} W.~P.~Wigner and  L.~Eisenbud, Phys.~Rev.~{\bf 72}~29(1947).
\bibitem{feshbach}  H. Feshbach, Ann. Phys. ~{\bf 5} 357 (1958).
\bibitem{mizuyama}  Kazuhito Mizuyama, Kazuyuki Ogata, Phys. Rev. C~{\bf 86}, 041603(R), 2012.
\bibitem{jost} R. ~Jost ~and ~A. ~Pais, ~Phys. ~Rev. ~\textbf{82}, ~840 ~(1951).
\bibitem{jost-hfb} K. Mizuyama, N. Nhu Le, T. Dieu Thuy, T. V. Nhan Hao, Phys. Rev. C {\bf 99}, 054607 (2019).
\bibitem{2nhalo}
  W.~Horiuchi~and~Y.~Suzuki,~Phys.~Rev.~C~{\bf 74},~034311~(2006).
\bibitem{2nhalo2}
  H.~Masui,~W.~Horiuchi,~and~M.~Kimura,~Phys.~Rev.~C~{\bf 101},041303(R).
\bibitem{dhalo}
  Masayuki~Matsuo, Kazuhito~Mizuyama, Yasuyoshi~Serizawa,~Phys.~Rev.~C~{\bf 71},~064326,~2005.
\bibitem{antihalo}
  K.~Bennaceur, J.~Dobaczewski,~and~M.~Ploszajczak, Phys.~Lett.~B {\bf 496},~154~(2000).
\bibitem{antihalo2}
  M.~Yamagami,~Eur.Phys.J.~A25~(2005)~569-570.
\bibitem{antihalo3}
  K.~Hagino~and~H.~Sagawa,~Phys.~Rev.~C~{\bf 84},~011303(R)~(2011).
\bibitem{pairrot}
  Nobuo~Hinohara1, and Witold~Nazarewicz,~Phys.~Rev.~Lett~{\bf 116},~152502 (2016)
\bibitem{2ntransfer}
  G.~Potel, A.~Idini, F.~Barranco, E.~Vigezzi and R.~A.~Broglia,
  Rep.~Prog.~Phys.~{\bf 76}~106301(2013).
\bibitem{kobayashi} Y.~Kobayashi, M.~Matsuo, Prog. Theor. Exp. Phys. 013D01 (2016).
\bibitem{jost-fano}
  K.~Mizuyama, N.~Nhu Le, and~T.~V.~Nhan Hao, Phys. Rev. C~{\bf 101}, 034601, 2020.
\bibitem{newton}
  Roger~G.~Newton,
  {\it Inverse Schrödinger Scattering in Three Dimensions},
  Springer, Berlin, Heidelberg.
  doi:https://doi.org/10.1007/978-3-642-83671-8
\bibitem{leonard}
  Leonard~S.~Rodberg and R.~M.~Thaler,
  {\it Introduction to the Quantum Theory of Scattering},
  Academic Press, New York, 1967.
\bibitem{shaperes}
  Amos~de~Shalit and Herman~Feshbach,
  {\it ``Theoretical Nuclear Physics: Nuclear structure''},
  John~Wiley~\&~Sons~Inc,~New~York,~page~87~(1974).
\bibitem{sasakawa}
  T.~Sasakawa, H.~Okuno, S.~Ishikawa,~and~T.~Sawada, Phys.~Rev.~C~\textbf{26},~42~(1982);
  ~T.~Sasakawa, Phys.~Rev.~C~\textbf{28},~439~(1983);
  ~T.~Sasakawa,~{\it ``Scattering theory''}~(Shokabo, Tokyo,1991),~ISBN978-4-7853-2321-9.
\bibitem{Mercer}
  J.~Mercer, Philosophical Transactions of the Royal Society A, 209 (441-458) 415-446, 1909.
  doi:10.1098/rsta.1909.0016
\end{thebibliography}
\end{document}